\documentstyle[11pt]{article}

\def\MSbar{\relax\ifmmode\overline{\rm MS}\else{$\overline{\rm MS}${ }}\fi}
\def \as{\relax\ifmmode a_s\else{$ a_s${ }}\fi}
\def\abar{\relax\ifmmode{\bar{a}}\else{$\bar{a}${ }}\fi}
\def\y{\relax\ifmmode{\bar{y}}\else{$\bar{y}${ }}\fi}
\def\x{\relax\ifmmode{\bar{x}}\else{$\bar{x}${ }}\fi}
\def\z{\relax\ifmmode{\bar{z}}\else{$\bar{z}${ }}\fi}
\def\albar{\relax\ifmmode{\bar{\alpha}}\else{$\bar{\alpha}${ }}\fi}
\def\albars{\relax\ifmmode{\bar{\alpha}_s}\else{$\bar{\alpha}_s${ }}\fi}
\def \asQ{\relax\ifmmode\bar  \alpha_s(Q)\else{$\bar \alpha_s(Q)${ }}\fi}
\def \asZ{\relax\ifmmode\bar  \alpha_s(M_Z)\else{$\bar \alpha_s(M_Z)${ }}\fi}
\def \asQm{\relax\ifmmode\bar \alpha_s(Q,m)\else{$\bar \alpha_s(Q,m)${ }}\fi}
\def \asQM{\relax\ifmmode\bar \alpha_s(Q,M)\else{$\bar \alpha_s(Q,M)${ }}\fi}

\newcommand{\nn}{\nonumber}
\newcommand{\la}{\label}      
\def\ie{\hbox{\it i.e.}{}}      
\def\eg{\hbox{\it e.g.}{}}      
\def\beqlab#1{\begin{equation}\label{#1}}
\newcommand{\Ds}{\displaystyle}

\newcommand{\bd}{\begin{displaymath}} \newcommand{\ed}{\end{displaymath}}
\newcommand{\ba}{\begin{eqnarray}}    \newcommand{\ea}{\end{eqnarray}}
\newcommand{\be}{\begin{equation}} \newcommand{\ee}{\end{equation}}
\newcommand{\baa}{\begin{array}{lll}} \newcommand{\eaa}{\end{array}}

\newcommand{\baz}{\begin{eqnarray*}}
\newcommand{\eaz}{\end{eqnarray*}}
\newcommand{\bb}{\begin{thebibliography}}
\newcommand{\eb}{\end{thebibliography}}

\begin{document}

\phantom{.}
\begin{center}
\large{RENORMALON CHAINS CONTRIBUTIONS TO THE NON-SINGLET \\
EVOLUTION KERNELS IN $[\varphi^3]_6$ AND QCD.}\\[0.5cm]

S.V. Mikhailov
\footnote{E-mail: mikhs@thsun1.jinr.dubna.su}\\[0.5cm]

{\it Joint Institute for Nuclear Research,
Bogoliubov  Lab. of Theoretical Physics,\\
141980, Moscow Region, Dubna, Russia}\\[0.5cm]

\end{center}
\begin{abstract}
The contributions to non-singlet evolution kernels $P(z)$, for the DGLAP
equation, and $V(x,y)$, for the Brodsky--Lepage equation, are
calculated for certain classes of diagrams including the renormalon
chains. Closed expressions are obtained for the sums of contributions
associated with these diagram classes. Calculations are performed in
the $[\varphi^3]_6$ model and QCD in the \MSbar scheme. The contribution
of one of the classes of diagrams dominates for a large number of flavours
$N_f \gg 1$. For the latter case, a simple solution to the Brodsky--Lepage
evolution equation is obtained.

\end{abstract}

\vspace {1cm}

PACS: 12.38.Cy, 12.38.-t, 13.60.Hb \\
Keywords: DGLAP evolution kernel, anomalous dimensions,
multiloop calculation
\section{Introduction}
Evolution equations play an important role in both inclusive \cite{L75}
and exclusive \cite{BL80} hard processes. They describe the dependence of
parton distribution functions of DIS and of the parton wave functions on
the renormalization parameter $\mu^2$.
The main ingredients of these equations are the perturbatively calculable
evolution kernels. The two-loop DGLAP evolution kernels $P(z)$ were
obtained in \cite{FLK81}, and a more complicated two-loop kernel $V(x,y)$
of the Brodsky--Lepage (BL) evolution equation  for meson wave functions
was calculated in \cite{DR84,MR85}. A three-loop calculation of these
kernels looks like being a tremendous problem. Recently, the results
of very complicated calculations of the first few elements of the
three-loop anomalous dimension of composite operators in
DIS -- $\gamma_{(2)}(N= 2, 4, 6, 8)$ have been presented in \cite{LRV94}.
Lately, these results were applied to improve the QCD analysis of
DIS experimental data \cite{KKSP96}.
Despite this important technical and phenomenological progress one
may feel insufficiency because of the absence of the whole kernel in
this order or, even at least, some parts of it. Of course, the kernel
admitting a physical probability interpretation is a more general and
convenient object from the theoretical point of view.

A method of estimating anomalous dimensions of composite operators
in the limit of a large number of flavours, $N_f$, has been suggested by J.
Gracey \cite{Gr94,Gr95}.
It is based on conformal properties of the theory at the critical point
 $g=g_c$ corresponding to the non-trivial zero
$g_c$ of the $D$-dimensional  $\beta$-function $\bar{\beta}(g_c)=0$. The
generating function has been constructed to obtain the leading large-$N_f$
asymptotics of the anomalous dimension $\gamma_{(n)}(N)$ to any order
$n$ of perturbation theory (PT).

In this letter, I propose a method of calculating certain classes of
multiloop diagrams directly for the DGLAP   kernel $P(z)$  as
well as for the BL kernel $V(x,y)$ in the \MSbar
scheme rather than their moments as in \cite{Gr94,Gr95}.
 These diagrams  contain insertion of chains of one-loop
self-energy parts (renormalon chains) into the one-loop diagrams
(see Fig.1 a,b) for kernels.
The kernels $P_{(n)}$ are obtained for any order $n$ of PT by analysing
of these ``dressed by chain" diagrams. Then, the kernel $P(z;g)$ ~($V$) for
diagrams with the totally dressed propagator
(\ie, in the ``all loops" approximation) is calculated; this kernel
appears to be a generating function to obtain the partial kernels $P_{(n)}$.
Moreover, this kernel $P(z;g)$ is the analytic function in variable
of a coupling constant $g$. The insertions of the renormalon chains lead
to converged PT series, and the usual renormalon singularities don't appear
in the renormgroup function calculations. This was found first for
$\beta$-  and  $\gamma$-functions in ~\cite{PalPas84}, the following
development of this approach is presented in ~\cite{David93}.
It is to be stressed that the method of obtaining $P$ or $V$ does not
depend on the nature of self-energy insertions and does not appeal to
the value of parameters
$N_f T_R, ~C_A/2$ or ~$C_F$ (for QCD case) associated with different
loops. Another distinctive feature is that the PT-improved evolution
kernels are calculated in the direct and standard way by using rather
elementary methods. This is possible due to a simple algebraic
structure of the counterterms of the diagrams considered.
In this letter, we mainly present technical results on resummation of
the above mentioned counterterms within
the framework of the $[\varphi^3]_6$ model and, in part, in QCD.
The solution to the BL evolution equation for a large number of flavors
$N_f$ is also briefly discussed.

To develop the diagrammatic analysis of multiloop evolution
kernels, let us first consider a toy model describing
$~\psi^*_i, ~\psi_i$, the charged ``quark" fields, and $\varphi$, the
``gluon" one, in $6$ space-time dimensions based on the Lagrangian
$\Ds L_{int}= g\sum^{N_f}_i (~\psi^*_i \psi_i \varphi)_{(6)}$.
The number of ``quark" flavors, $N_f$, will be considered
as an arbitrary free parameter associated with the ``quark" loops.
This theory has much in common with QCD:
 it is renormalizable and its $\beta$-function has a
structure similar to that of $\beta_{QCD}$,
$$
\beta(a) = -a^2b_0+\ldots, \quad b_0=\left(\frac{2}{3}-\frac{N_f}{6}\right),
\quad a=\frac{g^2}{(4\pi)^3}.
$$
Moreover, the one-loop evolution kernels $P_0$ and $V_0$, following from the
simplest triangular diagram in Figs.1a and 1b, turn out to be proportional
to the corresponding expressions for the same diagrams in QCD
\ba
 aP_0(z)=a(1-z), ~~~aV_0(x,y)=
a\left( \theta(y > x)\frac{x}{y}+\theta(\y > \x)\frac{\x}{\y}\right),
\ea
where \ $\y=1-y,~\x=1-x, \ldots$.
The similarity of the kernels in both the theories extends to the
two-loop level \cite{MR86}. The advantage of the $[\varphi^3]_6$
model is that the relevant Feynman integrals are simple,
and it is much easier to study the structure of multiloop
expressions.

In general, calculation of the evolution kernels $P(z; \alpha)$
\cite{FLK81} and $V(x,y; \alpha)$ \cite{MR85} is quite analogous to
that of the anomalous dimensions. The major modification is the change
$(kn)^N \to \delta(z - kn)$ of the vertex factor corresponding to the
composite operator, $k$ being the relevant momentum related to the incoming
``quark" line, see Figs. 1a,b. In these Figs., $p$ ( $\y p$ or $p$) is
the external momentum of the diagrams; $n$ is a light-like vector
($n^2=0$) introduced to pick out the symmetric traceless composite
operator; $pn=1$. Detailed examples of similar calculations can be found
in \cite{MR86}.

\section{First triangular diagrams for DGLAP evolution kernel}
In this section, the method will be demonstrated by using well-understood
 \cite{Gr95,PalPas84,MMS97} quark-loop insertions as an example.
Let us consider the one-loop triangular diagram
$\Gamma$ for kernels
in Fig.1c with the chain of self-energy part insertions $g_i$
into the gluon line.
Expressions for $P(z)$ or $V(x,y)$ in terms of the renormalization
constant $Z_{\Gamma}$ for the diagram $\Gamma$ in the \MSbar scheme
(see \cite{V80} and  \cite{MR85}) are given by
\ba
Z_{\Gamma}=1 - \hat{K}R'(\Gamma),
~~~~~P=-a\partial_a\left(Z^{(1)}_{\Gamma}\right) =
a\partial_a\left(\hat{K}_1 R' (\Gamma) \right).
\la{Z}
\ea
Here, $R'$ is the incomplete BPHZ $R$-operation; $\varepsilon=(6-D)/2$
and $D$ is the space-time dimension;
$\hat{K}$ picks out poles in $\varepsilon$, where $\hat{K}_1$ projects
a simple pole, and $Z^{(1)}$ is the
coefficient of the simple pole in the expansion of $Z_{\Gamma}$.
For $g_i$, being the one-loop insertions, expression
(\ref{Z}) can be simplified
following the definition of the $R'$-operation as
\ba
\Ds \hat{K}_1 R'(\Gamma)\equiv \hat{K}_1 R'(G\otimes \prod_i  g_i)=\hat{K}_1
\left[G\otimes \prod_i (1- \hat{K}_1) g_i \right]
\Rightarrow P_{(n)}=(n+1)a  \hat{K}_1
\left[G\otimes \prod_i^n  (1 - \hat{K}_1) g_i\right],
\la{Pn}
\ea
where $G$ denotes the ``outer triangle" part of the diagram $\Gamma$
without the chain of bubbles, and
\ba
\Ds Rg_i(k^2)& =& (1 - \hat{K}_1) g_i(k^2) = -\frac{aN_f}{\varepsilon}
\left(\gamma_{\varphi}(\varepsilon)
\left(\frac{\mu^2}{k^2}\right)^{\varepsilon}-\gamma_{\varphi}(0)\right)
\la{Rg}\\
\Ds \gamma_{\varphi}(\varepsilon) &=& C(\varepsilon)
B(2-\varepsilon, 2-\varepsilon).
\la{g}\ea
Here $B(a,b)$ is the Euler B-function,
$\gamma_{\varphi} = \gamma_{\varphi}(0)$ is the one-loop anomalous
dimension of the gluon field (at $N_f=1$), and
the constant $C(\varepsilon) = \Gamma(1-\varepsilon)\Gamma(1+\varepsilon)$
reflects the concrete choice of \MSbar scheme where
every loop integral is  multiplied by the scheme factor
$\Gamma(D/2-1)(\mu^2/4\pi)^{\varepsilon}$.
Note, the function $\gamma(\varepsilon)$ in (\ref{g}) will play an
important role in our consideration.
Substituting (\ref{Rg}) into (\ref{Pn}) and performing a direct
calculation (see, \eg, \cite{MR86}), one arrives at the expression
\ba \la{Pnz}
 P_{(n)}^{(1)}(z)=(n+1) a P_0(z) (-a N_f \gamma_{\varphi})^n \hat{K}_1
\left[\frac{C(\varepsilon)(1-\varepsilon)}{\varepsilon^{n+1}}
 z^{-\varepsilon} \sum_{j=0}^{n}
\left(\frac{\gamma_{\varphi}(\varepsilon)}{\gamma_{\varphi}(0)}\right)^j
\frac{\Gamma(1+(j+1)\varepsilon)}{\Gamma(1+j\varepsilon)\Gamma(1+\varepsilon)}
{n \choose j} \frac{(-)^{n-j}}{j+1}\right].
\ea
It is instructive to consider the properties of the first few
kernels $P_{(k)}^{(1)}(z)$. Their expressions are presented in Table 1 for
$k=1, \ldots ~8$. To obtain them, the FORM 2.0 program \cite{FORM} has
been used
\footnote{I am greatly indebted to
Dr. L. Avdeev who provided me with his brilliant
FORM-based program for expansion in $\varepsilon$, see \cite{Leo}. Note that
the contents of Table 1 is limited here only by place.}. The lessons
are : i) the perturbation theory for kernels $P_{(n)}^{(1)}(z)$ looks
like being improved in comparison with the
corresponding anomalous dimensions $\gamma_{(n)}(N)$. Indeed,
the term in $P_{(n)}^{(1)}(z)$ leading in $z$ is $\Ds \sim
\frac{\ln^n(z)}{n!}$, \ie, it is factorial suppressed in $n$,
while for the corresponding contribution to
$\gamma_n(N) \sim (n)^0$ is not suppressed; ii) one can see factorial
suppression in $n$ of all other logarithmic terms in $P_{(n)}^{(1)}(z)$;
iii) the $n$-bubble chain generates $\zeta(n)$  (the Riemann
 zeta-function) in the  non-logarithmic term in $P_{(n)}^{(1)}(z)$;
this may be traced back to the expansion of the Euler B-function
as their possible origin. These properties give hint about a possible
resummation of the $P_{(n)}^{(1)}(z)$-series. \\
Indeed, the closed expression for the sum of partial kernels
$\Ds P^{(1)}(z; A)= a\sum_{n=0}^{\infty}P_{(n)}^{(1)}(z)$ exist,
\ba \la{t1}
\Ds P^{(1)}(z; A) = a\sum_{n=0}^{\infty}P_{(n)}^{(1)}(z)
= a P_0(z) z^{-A}(1-A)
\left(\frac{\gamma_{\varphi}(0)}{\gamma_{\varphi}(A)}\right),
~\mbox{where} ~A=a N_f \gamma_{\varphi}
\ea
and the kernel $ P^{(1)}(z; A)$ is the generating function for
$P_{(n)}^{(1)}(z)$.
The proof of this assertion is presented in Appendix A.
This result possesses several remarkable properties:
\begin{itemize}
\item the $ P^{(1)}(z; A)$ becomes the dominant part of the
total perturbative kernel $P(z)$ when \\
$N_f \gg 1$. Below, the result for the kernel $P^{(1)}(z; A)$ will
be completed by taking into account similar corrections to the ``quark
leg" (see Fig.1d) whose contribution is proportional to $\delta(1-z)$.
\item  the analytic properties of $P^{(1)}(z; A)$ in the variable $A$ are
completely determined by the one-loop ``anomalous dimension"
$\gamma_{\varphi}(A)$ in $D$ dimensions (see Eqs.(\ref{Rg}, \ref{g})).
 The singularities of $P^{(1)}(z; A)$ correspond to
zeros of the function $B(2-A, 2-A)$ in $\gamma_{\varphi}(A)$. Of course,
$\gamma_{\varphi}(A)$ provides the scheme dependence in $P^{(1)}(z; A)$
due to the factor $C$ in the definition (\ref{g}).
\item the kernel $P^{(1)}(z; A)$ is an analytic function in variable $A$,
except for singularities at the points \\
$A=2+ k+1/2,$ $~k=0,~1,...$,
where the function has simple poles.
The nearest singularity appears at $A=5/2$, \ie, at $aN_f=15$,
determining the range of convergence of the PT series.
\end{itemize}

In a similar way, one can derive an expression for the sum of diagrams in
Fig.1 d connected with the anomalous dimension of the quark field
$\gamma_{\psi}^{(1)}(A)$ obtained in the ``all quark-loop" approximation.
Collecting the results of resummation in the main approximation in $A$,
which correspond to the diagrams in Figs.1c and d, we arrive at the final
expression
\ba \la{final2}
 P^{(1)}(z; A) - \delta(1-z)\gamma_{\psi}^{(1)}(A) =
 a\left(\frac{\gamma_{\varphi}(0)}{\gamma_{\varphi}(A)}\right)
 \left[\z z^{-A}(1-A) - \delta(1-z)\frac{(1-A)}{(3-A)(2-A)}\right].
\ea
Integration produces the following expression for the anomalous dimension
$\gamma(N, A)$ of the composite operator
\ba \la{gamma}
\gamma(N, A)&=& \int\limits_{0}^{1} z^N
\left(P^{(1)}(z; A) - \delta(1-z)\gamma_{\psi}(A)\right) dz
= \nn \\
&=&a\gamma_{\varphi}
 \frac{\Gamma(4-2A)}{\Gamma(2-A) \Gamma(1-A)^2\Gamma(1+A)}
 \left[\frac{\Gamma(N+1-A)}{\Gamma(N+3-A)}-
\frac{\Gamma(2-A)}{\Gamma(4-A)}\right].
\ea
Formula (\ref{gamma}) can be obtained by another method applied
in \cite{Gr94} to the QCD case (see Eq.(\ref{Mq1}) below).
Note that the anomalous dimension $\gamma(1, A)$ corresponds to
parton momentum. The equality $\gamma(1, A)=0$, following from the
r.h.s. of (\ref{gamma}), is expected; it signals that the parton momentum
is conserved.
\section{Other triangular diagrams contributions}
Consider now the class of diagrams in Fig. 2.
There are closed expressions for sums of  partial kernels
\ba  \la{t2a}
\Ds P^{(2a)}(z; B)&=&a \sum_{m=0}^{\infty}P^{(2)}_{(m)}(z)= a P_0(z)
\left((\z)^{-B}\frac{\gamma_{\psi}(0)}{\gamma_{\psi}(B)}\right),
\mbox{diag. in Fig.2a }, \\
\Ds \tilde{P}^{(2a)}(z;B)&=&a\sum_{m=0}^{\infty}(2-\delta_{0,n})
P^{(2)}_{(m)}(z) = a P_0(z) \left(2
(\z)^{-B}\frac{\gamma_{\psi}(0)}{\gamma_{\psi}(B)} -1 \right),
\mbox{diag. in Fig.2a} + MC, \la{t2b}\\
\Ds P^{(2b)}(z; B)&=&a \sum_{m=0}^{\infty} (m+1) P^{(2)}_{(m)}(z)=
 a P_0(z) \left(1+B\frac{d}{dB} \right)
\left((\z)^{-B} \frac{\gamma_{\psi}(0)}{\gamma_{\psi}(B)}\right),
~\mbox{diag. in Fig.2b }, \la{t2c} \\
 \delta(1-z) \gamma^{(2)}_{\psi} &=&
a \delta(1-z) \left(\frac{\gamma_{\psi}(0)}{\gamma_{\psi}(A)}\right)
  \frac{(1-A)}{(3-A)(2-A)},
~\mbox{diagr. in Fig.2c }; \la{t2d}
\ea
the functions $P(z;B)$ appear as generating functions for the
corresponding partial kernels.
Here $P^{(2)}_{(m)}(z)$ is the partial kernel with $m$ insertions into
one of the quark lines (see Fig.2a); $\gamma_{\psi}(\varepsilon)$ is
the one-loop anomalous dimension of the quark field in $D$-dimension;
for this model ~$\gamma_{\psi}(\varepsilon)=\gamma_{\varphi}(\varepsilon)$,
 ~$\gamma_{\psi} \equiv \gamma_{\psi}(0)= \gamma_{\varphi}(0)$,
 ~$B=a \gamma_{\psi}$; ~MC denotes a mirror-conjugate diagram.
The first Eq.(\ref{t2a}) corresponds to the diagrams in Fig. 2a where the
chain of quark self-energy parts is substituted only into the left quark
line of the triangle. To prove it, one has to repeat the way similar
to proof in Appendix A.
Equation (\ref{t2b}) corresponds to the sum of diagrams in Fig.2a
and its MC diagrams. This result will be used to restore the corresponding
kernel $V(x,y)$ in the next section. The analytic properties of the
functions $P^{(2a)}(z;B)$ and ~$\tilde{P}^{(2a)}(z;B)$ in the parameter $B$
are the same as for the kernel $P^{(1)}(z; B)$; they are determined by
the function $\gamma_{\psi}(B)$.
Equation (\ref{t2c}) corresponds to substitutions of the chains
into the both quark lines of the triangle in Fig.2b.
At least, Eq.(\ref{t2d}) corresponds to contributions to
the anomalous dimension of the quark field from the diagram in Fig.2c.

The contribution $P^{(2)}(z;B)$ will be suppressed in the parameter $B$
in comparison with $P^{(1)}(z;A)$, if $N_f \gg 1$, \ie, ~$A \gg B$.
Note, however, that the $N$-moments of the kernels $P^{(2a,b)}$
-- $\gamma^{(2)}(N,B)$ decrease in $N$ slowler than
$\gamma^{(1)}(N,A)$ corresponding to the  kernel $P^{(1)}$. Therefore, at
sufficiently large $N$ ~$\gamma^{(2)}(N,B) >   \gamma^{(1)}(N,A)$ for any
fixed parameters $A$ and $B$.

The expression for the kernel $P_{(n,m)}(z; A,B)$, corresponding to
insertions of different one-loop parts in both gluon
($n$--bubble insertions) and quark ($m$--self-energy part insertions)
lines of the triangular diagram (see Fig. 2d), is obtained as well.
This formula is similar to Eq.(\ref{Pnz}) for $P_{(n)}$,
but looks more cumbersome and is not shown here.
The partial kernels $P_{(n,m)}(z; A,B)$ can be obtained by using the FORM
program, in general, for any given $n$ and $m$. For illustration, we
demonstrate here the first nontrivial kernel $aP_{(1,1)}$
$$
P_{(1,1)} = P_0(z) AB
\left( \frac1{2}\left[\ln(z)+\frac{8}{3}\right]^2 -
\frac{20}{9} + \frac1{2}\left[\ln(\z)+\frac{5}{3}\right]^2
- \frac{31}{18}
-2 \left[\ln(z)+\frac{8}{3}\right] \left[\ln(\z)+\frac{5}{3}\right]
+\frac{80}{9} -2 \zeta(2) \right)
$$
in comparison with partial kernels of the same order in $a$,
$P_{(2,0)}$
~($\equiv P_{(2)}^{(1)}$) following from the expression for $P^{(1)}$
 and $P_{(0,2)}$ ~($\equiv 3P^{(2)}_{(2)}$), from $P^{(2b)}$
$$
P_{(2,0)} = P_0(z) A^2
\left( \frac1{2!}\left[\ln(z)+\frac{8}{3}\right]^2 - \frac{20}{9}\right),
~~~P_{(0,2)} =3P_0(z) B^2
\left( \frac1{2!}\left[\ln(\z)+\frac{5}{3}\right]^2 - \frac{31}{18}\right).
$$
To complete the section, we conclude that contributions to kernel
$P$ from any one-loop insertions into the lines of the triangular
diagram are available now.

\section{Triangular diagrams for the Brodsky--Lepage evolution kernel}
Here, some partial results of the bubble resummation for the BL kernels
$V$ are presented. We obtain them as a ``byproduct" of our previous
results for the kernel DGLAP $P$. The results obtained may be used,
in particular, to check the regular calculation of the kernel $V$ in high
orders of PT.

Note, the diagrams for the BL kernel differ from the DGLAP diagrams only
by the ``exclusive" kinematics of the input momentum, compare the diagrams
in Fig.1a and 1b. So, one can repeat the proof similar to that
in Appendix A for this case.
There is another far more elegant way -- to use exact
relations between $V$ and $P$ kernels for triangular diagrams
which were established for any order of perturbation theory in \cite{MR85}.
These relations work for both the $[\varphi^3]_6$ model and QCD.
I quote these propositions without proofs.

Let the diagram in Fig.1c have a contribution to the DGLAP kernel
in the form $P(z)= p(z) +$ $\delta(1-z) \cdot C$;
then its contribution to the BL kernel is
\ba \la{V1}
\Ds V(x,y) = {\cal C} \theta(y > x) \int^{\frac{x}{y}}_0
\frac{p(z)}{\bar z} dz + \delta(y-x) \cdot C,
\ea
where $ {\cal C} \equiv 1 + \left(x \to \bar x, y \to \bar y \right)$.
Substituting Eq. (\ref{final2}) for $P^{(1)}(z)$
into relation (\ref{V1}) we
immediately derive the expression for $V^{(1)}(x,y)$
\ba \la{V1a}
\Ds V^{(1)}(x,y; A) =  {\cal C} a
\left(\frac{\gamma_{\varphi}(0)}{\gamma_{\varphi}(A)}\right)
\left[ \theta(y > x)\left(\frac{x}{y}\right)^{1-A}
 - \frac{1}{2}\delta(y-x)
\frac{(1-A)}{(3-A)(2-A)} \right].
\ea
The later expression may independently be verified by other relations
by reducing any $V$ to $P$ \cite{MR85,DMRGH88} \\
($V \to P$ reduction)
\ba \la{reduct}
\Ds V^{(1)}(x,y; A) =  {\cal C}\theta(y > x)F(\frac{x}{y};A) -
\delta(y-x) \cdot C(A)
\to P^{(1)}(z; A) && \nonumber \\
\mbox{where} ~~P^{(1)}(z; A)= \theta(1-z) \z \frac{d}{dz} F(z;A)-
\delta(1-z) \cdot C,
&&
\ea
indeed, substituting (\ref{V1a}) into (\ref{reduct})
we return to the same Eq.(\ref{final2}) for $P^{(1)}(z)$. Moreover,
the first term of the Taylor expansion of $V^{(1)}(x,y; A)$ in $A$ coincides
with the results of the two-loop calculation in \cite{MR86}.
The contribution $V^{(1)}$ should dominate for $N_f \gg 1$ in the whole
kernel $V$, and moreover, the function $V^{(1)}(x,y)$ possesses an important
symmetry.
Really, the function ${\cal V}(x,y; A)=V^{(1)}(x,y; A) \cdot (\y y)^{1-A}$
is symmetric under the change $x \leftrightarrow y$,
${\cal V}(x,y)={\cal V}(y,x)$. This symmetry allows us to obtain the
eigenfunctions $\psi_n(x)$ of the equation
\ba \la{ev}
&& \int\limits_{0}^{1} V^{(1)}(x,y; A)\psi_n(y) dy= \gamma(n;A) \psi_n(x), \\
&&\Ds \psi_n(y) \sim (\y y)^{d_{\psi}(A) - 1}
C_{n}^{(d_{\psi}(A) - \frac1{2})} (y-\y),
~\mbox{here}
~~d_{\psi}(A) = D_A/2-1, ~~D_A=6-2A, \la{solv1}
\ea
$d_{\psi}(A)$ is the effective dimension of the scalar field when the
anomalous dimension is taking into account, and $C_{n}^{(\alpha)}(z)$
are the Gegenbauer polynomials of an order
of $\alpha$. This form of eigenfunctions $\{\psi_n(y)\}$ as well as the
 $x \leftrightarrow y$ symmetry of the function ${\cal V}(x,y)$ are the
consequences of conformal symmetry conservation (see \cite{Mak81} for
details) for the sum of diagrams
under consideration \footnote{ When this letter was considered
by the Editorial Board, the work \cite{GK97} appeared, where a similar
solution has been presentned for the case of QCD evolution kernel.}.
Equation (\ref{ev}) is tightly connected with the BL evolution equation
in a special case when the $\beta$-function $\beta=0$.
In this case, the BL equation is simplified, the variables are separated,
and the evolution equation may be reduced to (\ref{ev}). The partial
solution to the BL
equation turns out to be proportional to $\psi_n(y)$ in (\ref{solv1}).

For the diagrams in Fig. 2a plus its MC diagrams, the relation
between $\tilde{V}^{(2a)}$ and $\tilde{P}^{(2a)}$ is (see \cite{MR85})
\ba \la{V2}
\Ds \tilde{V}^{(2a)}(x,y) =  {\cal C}\theta(y > x) \frac1{2}
\left[\frac1{\bar y}\tilde{P}^{(2a)}(x)+\frac1{y}\tilde{P}^{(2a)}(\bar x)
- \frac1{\y}\tilde{P}^{(2a)}\left(\frac{x}{y}\right)\right].
\ea
Substituting (\ref{t2b}) into (\ref{V2}) one arrives at the expression for
$\tilde{V}^{(2a)}(x,y)$
\ba
\Ds \tilde{V}^{(2a)}(x,y) =  {\cal C}  a\theta(y > x)
\left\{
\left[ \frac{\x^{1-B}}{\y}+\frac{x^{1-B}}{y} - \frac{1}{\y}
\Bigl(1-\frac{x}{y} \Bigr)^{1-B} \right]
\left(\frac{\gamma_{\psi}(0)}{\gamma_{\psi}(B)}\right)
- \frac{x}{y} \right\},
\ea
which satisfies the same check as in the previous case.
Of course, this formula does not cover the diagrams in Fig. 2b, where
both the quark lines are dressed.

\section{ The results for QCD evolution kernels}
The assertions about diagram contributions to $P$ and $V$, similar to
the presented above, are also valid for the QCD diagrams.
Their proof does not contain essentially new elements,
but looks rather cumbersome.
The complete results for the QCD evolution kernels will be presented in
a subsequent paper. Here, at first, the QCD results for triangular diagrams
in Fig. 1 c,d in the Feynman gauge are discussed.
Based on the proof in Appendix A in the QCD version, one can derive the
result for the sum of diagrams, in Fig. 1c, in QCD

\ba \la{Pqcd}
P^{(1c)}(z; A) = \as C_F 2\z \cdot (1-A)^2 \frac{\Pi(0)}{\Pi(A)} z^{-A} -
 \as C_F\cdot \delta(1-z)
 \left(\frac{\Pi(0)}{\Pi(A)}\frac1{(1-A)} -1 \right).
\ea
Here, $\Ds \as = \frac{\alpha_s}{4\pi}$; ~$\Pi(\varepsilon)$ is twice
the contribution to a one-loop
$D$--dimensional anomalous dimension of the gluon field and
$\varepsilon=(4-D)/2$; $\Pi(0)$ is the contribution to the standard
anomalous dimension; $A=-\as\Pi(0)$.
It should be emphasized again that the form of Exp.(\ref{Pqcd}) does not
depend on the nature of self-energy insertions into the gluon line. One
can use it for the resummation of both the quark $(\sim N_f T_R)$ and
gluon ($\sim C_A/2$) loops.
Substituting into the general formula (\ref{Pqcd}), the well-known
expressions for $\Pi(\varepsilon)$ from the quark or gluon (the ghost
loop is also added) loops
\ba \la{Piq}
\Pi_q(\varepsilon) &=& -8 N_f T_R B(D/2,D/2) C(\varepsilon),\\
\Pi_g(\varepsilon)& =&  \frac{C_A}{2} B(D/2-1,D/2-1)
\left(\frac{3D-2}{D-1}\right) C(\varepsilon), \la{Pig}
\ea
one obtains the expression $P^{(1)}_q(z; \delta)$ for the quark--loop
insertions
\ba \la{Pq}
P^{(1c)}_q(z; \delta)& =& \as C_F 2\z  z^{-\delta}
\frac{ (D_q/2-1)^2 B(2,2)}{B(D_q/2,D_q/2)C(\delta)} \nn \\
&&- \as C_F\delta(1-z)
\left(\frac{B(2,2)}{B(D_q/2,D_q/2)C(\delta)}\frac1{(D_q/2-1)} -1 \right)
\ea
 where
 $ ~D_q=4-2\delta$, ~$\Ds \delta= -\as \Pi_q(0) = \as N_f T_R \frac{4}{3}$,
~$C(\delta)$ is the scheme factor, \\
and the expression $P^{(1c)}_g(z; \epsilon)$ for the gluon--loop insertions
\ba \la{Pg}
P^{(1c)}_g(z; \epsilon) &=& \as C_F 2\z z^{-\epsilon}
\frac{10}{3}\frac{(D_g/2-1)^2(D_g-1)}{(3D_g-2)B(D_g/2-1,D_g/2-1)C(\epsilon)}
 \nn  \\
&& - \as C_F\delta(1-z)
\left(\frac{10}{3}\frac{(D_g-1)}{(3D_g-2)B(D_g/2-1,D_g/2-1)C(\epsilon)}
\frac1{(D_g/2-1)}-1 \right),
\ea
where $\Ds ~D_g=4-2\epsilon, ~\epsilon = -\as \Pi_g(0)=-\as C_A \frac{5}{3}$.
The expression (\ref{Pg}) for $P^{(1c)}_{g}(z; \epsilon)$ has no any
singularity in the parameter $\Ds \epsilon < 0$ due to the asymptotic
freedom.

Consider Eq.(\ref{Pq}) for $P^{(1c)}_{q}(z;\delta)$ in detail.
By adding the contributions from the diagrams in Fig.1d to (\ref{Pq}),
one can find that the second term cancels in part and the final
expression turns out to be the expected ``plus form" (see, \eg
~\cite{FLK81})
\ba \la{Pq1}
P^{(1c,d)}_q(z; \delta) = \as C_F 2
\left( \z  z^{-\delta}
\frac{ (D_q/2-1)^2 \Pi_q(0)}{\Pi_q(\delta)} \right)_{+},
~\mbox{\ie,} ~\int^{1}_{0} P^{(1c,d)}_q(z; \delta)dz=
\gamma_{q}^{(1c,d)}(N=0, \delta)=0
\ea
due to the current conservation. The analytic properties of
$P^{(1c,d)}_q(z; \delta)$ in $\delta$ are the same as for its scalar
analogy $P^{(1)}(z; A)$ (see (\ref{t1}) in sect.2); they are determined
by the behavior of the function $\Pi_q(\delta)$ in $\delta$, see
Eq.(\ref{Piq}). The nearest singularity of $P^{(1)}_q(z; \delta)$
in $\delta$ appears at $\as N_f = 15/4$. The moments of Eq. (\ref{Pq1})
agree with the corresponding part of the generating function in \cite{Gr94}
\ba \la{Mq1}
\gamma_{q}^{(1c,d)}(N, \delta)= -\as C_F \frac{2}{3}
\frac{N(D_q+N-1)}{(D_q/2+N)(D_q/2-1+N)}
   \left[ \frac{(D_q/2-1)}{B(D_q/2,D_q/2)C(\delta)D_q} \right],
\ea
see the first term in Eq.(14) in \cite{Gr94}, and ref. \cite{MMS97}
(note that our moments differ in sign from the definition of the
anomalous dimension of composite operators used there).

To complete the QCD calculations of the kernel $P^{(1)}$, we need the
contribution from the last diagram in Fig.1e with the
chain in the gluon line that is inserted into the composite
operator. It can be obtained in a similar way, as in the previous
QCD--calculations for Eq.(\ref{Pqcd})
\ba \la{Pqcd2}
P^{(1e)}(z; A) = \as C_F 2 \cdot \left(\frac{2 z^{1-A}}{1-z}
\frac{\Pi(0)}{\Pi(A)} \right)_{+},
\ea
and the expression automatically has the ``plus form", see
~\cite{MR85} for details.
Collecting Eq.(\ref{Pq1}) and Eq.(\ref{Pqcd2}), one easily
arrives at the complete QCD expression for $P_q^{(1)}(z; \delta)$
%with the main quark--loop insertions
\ba \la{final}
P_q^{(1)}(z; \delta) =  \as C_F 2 \cdot
\left[\z z^{-\delta}(1-\delta)^2 + \frac{2 z^{1-\delta}}{1-z} \right]_{+}
\frac{\Pi_q(0)}{\Pi_q(\delta)}.
\ea
The contribution $P_q^{(1)}(z; \delta)$ is gauge invariant.
The moments of $P_q^{(1)}(z; \delta)$ again  agree with the complete
generating function obtained in \cite{Gr94}  (see Eq.(14) there).
To obtain the contribution from both kinds of insertions into the
gluon line one would take certain substitutions in (\ref{final}):
$\Pi_q(\varepsilon)+ \Pi_g(\varepsilon) \to \Pi_q(\delta)$ and
$\varepsilon \to \delta$, here $\varepsilon =-a (\Pi_q(0)+ \Pi_g(0))$.
Of course, for this case a final result must be gauge dependent.

\section{ Conclusion}
The method for calculating certain classes of multiloop diagrams
for the kernel $P(z)$ of the non-singlet DGLAP evolution equation
is presented. These multiloop diagrams appear due to the insertion of
chains of one-loop self-energy parts (renormalon chains) into the lines
of the one-loop diagrams for the kernel. Closed expressions
$P^{(1,2)}(z, a_{1,2})$ are found for sums of all the diagrams
which belong to two of the diagram classes,
see sect. 2-3. The corresponding proofs are based on a simple algebraic
structure of the counterterms for the diagrams under consideration;
this is outlined in Appendix A. Moreover, the kernels $P^{(i)}(z, a_i)$
are generating functions for the partial kernels $P_{(n)}^{(i)}(z)$
which correspond to the $n$-bubble insertion. The contribution
$P^{(1)}(z; a_1)$ (corresponding to one of the diagram classes) would
dominate in the kernel for $N_f \gg 1$. The analytic properties of the
function $P^{(i)}(z, a_i)$ in the variable $a_i$ are briefly discussed.
The expressions for partial kernels $P_{(n,m)}(z;a_1,a_2)$ for the
diagrams of a ``mixed class", in any order of perturbation theory can also
be obtained by using the FORM program.

The contributions $V^{(i)}(x,y; a_i)$ to the Brodsky--Lepage kernel are
obtained in sect. 4 for the same classes of diagrams as a ``byproduct" of
the previous technique. When $N_f \gg 1$, a special solution to the
Brodsky--Lepage equation is derived. We emphasize that the method of
calculating the evolution kernels $P^{(i)}$ or $V^{(i)}$ does not depend
on the nature of self-energy insertions and does not appeal to
the value of parameters $N_f T_R, ~C_A/2$ or ~$C_F$ (for QCD case)
associated with different loops.

The method and results are exemplified with a simple
$[\varphi^3]_6$ model.
It is clear that all the results obtained above in the framework of the
scalar model have a wider meaning and can be applied to the QCD case.
Here, in sect. 5, some QCD results are presented too; in particular, the
kernel $P_{q}^{(1)}(z; \delta)$, that corresponds to
the diagram dressed by the main quark-loop chain, is derived.
The anomalous dimension $\gamma_{q}^{(1)}(N, \delta)$
corresponding to this kernel agrees with the generating function
obtained earlier by Gracey \cite{Gr94,Gr95}. The contribution
$P^{(1)}_g(z; \epsilon)$ from the diagrams with the gluon-loop
chain is derived in the same way.
The same can be obtained for the mixed case when both kinds of bubbles
(quark and gluon) are involved.
The completed QCD case will be considered in detail in a subsequent paper.

\vspace{3mm}

\centerline{\bf Acknowledgements}
\vspace{2mm}
The author is grateful to Dr. L. Avdeev, Dr. M. Kalmykov  and Dr. A.
Bakulev for their help in programming the calculations by FORM,
Dr. A. Grozin and Dr. R. Ruskov for
fruitful discussions of the results, to Dr. N. Stefanis, Dr. G. Korchemsky
and the referee of this letter for careful reading of the manuscript and
useful remarks.
This investigation has been supported in
part by the Russian Foundation for Fundamental Research (RFFR)
96-02-17631  and INTAS 93-1180 ext
\begin{appendix}
\appendix
\section{Appendix}
\renewcommand\theequation{\thesection{.\arabic{equation}}}
\setcounter{equation}{0}
Here, formular (\ref{t1}) is proved; all others resummation formulae
in the text above can be obtained in a similar manner. \\
{\bf Proof}. One could split the sum in square brackets, Eq.(\ref{Pnz}),
 into two parts
\ba \la{qt1}
&&(n+1) \sum_{j=0}^{n}
\left( F(\varepsilon)\right)^j
\frac{\Gamma(1+(j+1)\varepsilon)}{\Gamma(1+j\varepsilon)\Gamma(1+\varepsilon)}
{n \choose j} \frac{(-)^{n-j}}{j+1} = S_{(n+1)}(\varepsilon)
+ \frac{(-)^{n}}{F(\varepsilon)C(\varepsilon)}, \\
&&\mbox{where} ~~S_{(n+1)}(\varepsilon) \equiv
\left[ \sum_{j=0}^{n+1} \left( F(\varepsilon) \right)^{j - 1}
\frac{\Gamma(1+j \varepsilon) (-)^{n+1-j}}{\Gamma(1+(j-1)\varepsilon)
\Gamma(1+\varepsilon)}{n+1 \choose j} \right].
 \la{Sn}\ea
Here, we use the notation:
$\Ds F(\varepsilon) \equiv
\frac{\gamma_{\varphi}(\varepsilon)}{\gamma_{\varphi}(0)}$,
 ~$S_{(n+1)}$ is the sum including ``all combinatoric factors"
of the l.h.s. of (\ref{qt1}), and the last term in the r.h.s. of
(\ref{qt1}) may be associated with the $j=-1$ term in sum.
The term $S_{(n+1)}$ in the r.h.s. of (\ref{qt1})
does not contribute to the pole part, by virtue of Lemma:
$\Ds S_{(n+1)}(\varepsilon) \leq O(\varepsilon^{n+1})$.
The expression for $P_{(n)}^{(1)}(z)$ can be derived
by substituting decomposition (\ref{qt1}) into (\ref{Pnz}) and using
the above mentioned Lemma
\ba
P_{(n)}^{(1)}(z)
&=& a P_0(z) (-A)^n \hat{K}_1
\left[\frac{1}{\varepsilon^{n+1}} z^{-\varepsilon}(1-\varepsilon)
\left( C(\varepsilon) S_{(n+1)}(\varepsilon)
+ \frac{(-)^n}{ F(\varepsilon)}\right) \right]= \nn \\
&=& a P_0(z) (A)^n \hat{K}_1
\left[\frac{1}{\varepsilon^{n+1}}  z^{-\varepsilon}
\frac{(1-\varepsilon)}{F(\varepsilon)}\right]
= a P_0(z) \frac{(A)^n}{n!}
\left. \left\{ \frac{d^{n}}{d\varepsilon^{n}}
\left[ z^{-\varepsilon} \frac{(1-\varepsilon)}{F(\varepsilon)}\right]
\right\}\right|_{\varepsilon=0}.
\la{Pnz1}\ea
Now, let us take a sum of $P_{(n)}^{(1)}(z)$ over all $n$ using
Exp.(\ref{Pnz1}); the differentials are naturally collected into the
``shift argument operator" $\exp\left(A \partial_{\varepsilon}\right)$
\ba \la{shift}
\Ds  P^{(1)}(z; A) = a \sum_{n} P_{(n)}^{(1)}(z) =
aP_{0}(z) \left\{ \exp\left(A \partial_{\varepsilon}\right)
\left. \left[ z^{-\varepsilon} \frac{(1-\varepsilon)
\gamma_{\varphi}(0)}{\gamma_{\varphi}(\varepsilon)} \right]
\right|_{\varepsilon=0}
\right\}.
\ea
Finally, performing the shift of argument in the squar bracket in
(\ref{shift}), we arrive at the expression
\ba \la{Pta}
\Ds P^{(1)}(z; A)=a\sum_{n=0}^{\infty} P_{(n)}^{(1)}(z) = a P_0(z)
z^{-A}(1-A) \left(\frac{\gamma_{\varphi}(0)}{\gamma_{\varphi}(A)} \right),
\ea
and the partial kernels $P_{(n)}^{(1)}(z)$ appear in the Taylor
expansion of $P^{(1)}(z; A)$ in $A$ by construction
\rule{6pt}{6pt} \\
Lemma: ~$S_{(n+1)}(\varepsilon) \leq O(\varepsilon^{n+1})$
~for any analytic function $F(\varepsilon)$ at point $\varepsilon = 0$.

Let us consider the expansion of every element of the sum in $S_{(n+1)}$
\be \la{lem1}
\frac{F(\varepsilon)^j}{F(\varepsilon)}
\frac{\Gamma(1+j \varepsilon)}{\Gamma(1+(j-1)\varepsilon)
\Gamma(1+\varepsilon)}
\ee
in powers of $\varepsilon$ up to $\varepsilon^n$.
Any power $\varepsilon^m$ of this expansion is accompanied by
 powers $j^l$, where $l \le m$, and a coefficient that does not depend
on $j$.
Therefore, the expansion of $S_{(n+1)}(\varepsilon)$ as a whole in the
power series $\varepsilon^m$ will generate coefficients of the powers
which are composed only of the elements proportional to
$\Ds \sum_{j=0}^{n+1}j^l (-)^{n+1-j} {n+1 \choose j}$.
All these elements are equal to $0$ for $m \leq n$ in virtue of the identity
\be \label{A7}
 \sum_{j=0}^{n+1}j^l (-)^{n+1-j} {n+1 \choose j} = 0,
 ~~\mbox{if} ~~l < n+1.
\ee
Therefore, the power expansion of $S_{(n+1)}(\varepsilon)$ can be started
with the power higher than $\varepsilon^{n}$. Note, the similar trick with
the identity (\ref{A7}) had been used in \cite{PalPas84}.
The proof implies an obvious generalization of the elements (\ref{lem1})
that can be constructed as superpositions of $\Gamma$-functions
depending on $j$
only through the arguments like $1+j \varepsilon , ~1+(j-1) \varepsilon,
...$. \rule{6pt}{6pt}

\end{appendix}

\newpage
{ {
\begin{tabular}{|l|l|} \hline
$n$ & The partial kernels $\Ds P_{(n)}^{(1)}(z)$ ~~(the common factor
~$aP_0(z) A^n$ is dropped)\\ \hline \hline
 1  &{\strut$\Ds \frac{\left[\ln(z)+\frac83\right]^1}{1!}_{\vphantom{\vbox to
 7mm{}}}$\vphantom{\vbox to 7mm{}} } \\ \hline
 2  &{\strut$\Ds \frac{\left[\ln(z)+\frac83\right]^2}{2!}-\frac{20}{9}_{
 \vphantom{\vbox to 10mm{}}}$\vphantom{\vbox to 10mm{}} } \\ \hline
 3  &{\strut$\Ds \frac{\left[\ln(z)+\frac83\right]^3}{3!}-\frac{20}9
\frac{\ln(z)}{1!}- \left(\frac{256}{81}+2\zeta(3)\right)_{\vphantom{\vbox to
 10mm{}}}$\vphantom{\vbox to 10mm{}} } \\ \hline
 4  &{\strut$\Ds \frac{\left[\ln(z)+\frac83\right]^4}{4!}-\frac{20}9\frac{
 \ln^2(z)}{2!}-
  \left(\frac{256}{81}+2\zeta(3)\right)\frac{\ln^1(z)}{1!}-\frac{512}{243}
  +3\zeta(4)-\frac{16}{3}\zeta(3)_{\vphantom{\vbox to 7mm{}}}$\vphantom{\vbox
   to 7mm{}} } \\ \hline
 5  &{\strut$\Ds \frac{\left[\ln(z)+\frac83\right]^5}{5!}-\frac{20}9
  \frac{\ln^3(z)}{3!}-
  \left(\frac{256}{81}+2\zeta(3)\right)\frac{\ln^2(z)}{2!}-
  \left(\frac{512}{243}
 +3\zeta(4)-\frac{16}{3}\zeta(3)\right)\frac{\ln^1(z)}{1!}
 -\frac{4096}{3645}$} \\
  &{$\Ds +8\zeta(4)-6\zeta(5)-\frac8{3}_{\vphantom{\vbox to 7mm{}}}$
  \vphantom{\vbox to 7mm{}} } \\ \hline
 6&{\strut$\Ds \frac{\left[\ln(z)+\frac83\right]^6}{6!}-\frac{20}9\frac{
  \ln^4(z)}{4!}-
  \left(\frac{256}{81}+2\zeta(3)\right)\frac{\ln^3(z)}{3!}-
  \left(\frac{512}{243}
  +3\zeta(4)-\frac{16}{3}\zeta(3)\right)\frac{\ln^2(z)}{2!}+
  \left(-\frac{4096}{3645} \right. $}\\
  &{$\Ds \left. +8\zeta(4)-6\zeta(5)-\frac8{3}\right)
  \frac{\ln^1(z)}{1!}-\frac{16384}{32805}+
  10\zeta(6)-16\zeta(5)+4\zeta(4)+2\zeta^2(3)_{\vphantom{\vbox to 7mm{}}}$
  \vphantom{\vbox to 7mm{}} } \\ \hline
 7&{\strut$\Ds \frac{\left[\ln(z)+\frac83\right]^7}{7!}-\frac{20}9
 \frac{\ln^5(z)}{5!}-
  \left(\frac{256}{81}+2\zeta(3)\right)\frac{\ln^4(z)}{4!}-
  \left(\frac{512}{243}
  +3\zeta(4)-\frac{16}{3}\zeta(3)\right)\frac{\ln^3(z)}{3!}+
  \left(-\frac{4096}{3645} \right.$}\\
  &{$\Ds \left.+ 8\zeta(4)-6\zeta(5)-\frac8{3}\right)
  \frac{\ln^2(z)}{2!} +\left(-\frac{16384}{32805}+
  10\zeta(6)-16\zeta(5)+4\zeta(4)+2\zeta^2(3)\right)
  \frac{\ln^1(z)}{1!}-18\zeta(7)$} \\
  &{$\Ds +\frac{80}{3}\zeta(6) -8\zeta(5)
  -6\zeta(3)\zeta(4)+\frac{16}{3}\zeta^2(3)_{\vphantom{\vbox to 7mm{}}}$
  \vphantom{\vbox to 7mm{}} } \\ \hline
 8&{\strut$\Ds \frac{\left[\ln(z)+\frac83\right]^8}{8!}-\frac{20}9
  \frac{\ln^6(z)}{6!}-
  \left(\frac{256}{81}+2\zeta(3)\right)\frac{\ln^5(z)}{5!}-
  \left(\frac{512}{243}
  +3\zeta(4)-\frac{16}{3}\zeta(3)\right)\frac{\ln^4(z)}{4!}+
  \left(-\frac{4096}{3645} \right.$}\\
  &{$\Ds \left.+ 8\zeta(4)-6\zeta(5)-\frac8{3}\right)
  \frac{\ln^3(z)}{3!} +\left(-\frac{16384}{32805}+
  10\zeta(6)-16\zeta(5)+4\zeta(4)+2\zeta^2(3)\right)
  \frac{\ln^2(z)}{2!}-\left( 18\zeta(7) \right.$} \\
  &{$\Ds \left.+\frac{80}{3}\zeta(6) -8\zeta(5)
  -6\zeta(3)\zeta(4)+\frac{16}{3}\zeta^2(3) \right)\ln(z)
  +\frac{63}{2}\zeta(8)-48\zeta(7)+\frac{40}{3}\zeta(6)+\frac{9}{2}
  \zeta^2(4)$} \\
  &{$\Ds+12\zeta(3)\zeta(5)-16\zeta(3)\zeta(4)
  +\frac{8}{3}\zeta^2(3)_{\vphantom{\vbox to 7mm{}}}$
  \vphantom{\vbox to 7mm{}} } \\ \hline
 \end{tabular}
 }} \\
\vspace{1cm}

{\bf Table 1}. The results of the $P_{(n)}^{(1)}(z; A)$ calculations,
the $\zeta(n)$ is the Riemann zeta--function; note that  $\zeta(2)$
and the Euler constant $\gamma_E$  do not appear in this expansion.

\begin{thebibliography}{99}
\bibitem{L75}
   V.N. Gribov and L.N.~Lipatov, Sov.J. Nucl.Phys.~{\bf 15} (1972) 438; 675;
              L.N.~Lipatov, Sov.J. Nucl. Phys.~{\bf 20} (1975) 94;
              Y.L.~Dokshitser, JETP ~{\bf 46} (1977) 641;
              G. Altarelli and G. Parisi, Nucl. Phys.~{\bf B 126} (1977) 298.
\bibitem{BL80} S.J. Brodsky, and G.P. Lepage,  Phys. Lett.~{\bf B 87} (1979)
359; Phys. Rev.~{\bf D 22} (1980) 2157.
\bibitem{FLK81} E.G. Floratos, R. Lacaze and C.Kounnas,  Phys. Lett.
~{\bf B 98} (1981) 89; 285.

\bibitem{DR84} F.M. ~Dittes and A.V.~Radyushkin, Phys. Lett.~{\bf B 134} (1984)
 359; ~M.H. ~Sarmadi,  Phys. Lett.~{\bf B 143} (1984) 471;
  ~S.V. Mikhailov and A.V.Radyushkin,
{\em ``Evolution kernel for the pion wave function: two loop QCD calculation
in Feynman gauge."}.
Dubna preprint JINR P2-83-721 (1983).

\bibitem{MR85} S.V. Mikhailov and A.V.Radyushkin, Nucl. Phys.~{\bf B 254}
 (1985) 89.

\bibitem{LRV94}
S.A.~Larin,  T.~van Ritbergen, J.A.M.~Vermaseren, Nucl.Phys.~{\bf B 427}
(1994) 41;\\
 S.A.~Larin, P.~Nogueira,T.~van Ritbergen, J.A.M.~Vermaseren,
{\em The three loop QCD calculation of the moments of deep inelastic
structure functions.}
NIKHEF-96-010, hep-ph/9605317

\bibitem{KKSP96}
G.~Parente, A.V.~Kotikov, V.G.~Krivokhizhin, Phys. Lett.~{\bf B 333} (1994)
190; A. L. ~Kataev, A. V.~Kotikov, G. Parente and A.V.~Sidorov,
Phys. Lett.~{\bf B 388} (1996) 179;

\bibitem{Gr94}
J.~A.~Gracey,  Phys. Lett.~{\bf B 322} (1994) 141;
\bibitem{Gr95}
J.~A.~Gracey, {\em Renormalization group functions of QCD in
large-$N_f$}, talk presented at Third International Conference on
the Renormalization Group, JINR, Dubna, Russia, 26-31 August, 1996,
hep-th/9609164.

\bibitem{PalPas84} A. Palanques-Mestre and P. Pascual,
Comm. Math. Phys. 95 (1984) 277.

\bibitem{David93}
D.~J. Broadhurst, Z. Phys. {\bf C} 58, 339 (1993); D.~J. Broadhurst and
A.~G.~Grozin, Phys. Rev.~{\bf D 52} (1995) 4082;
~M.~Beneke and V.~M. ~Braun,
Nucl. Phys. ~{\bf B 426} (1994) 301.

\bibitem{MR86} S.V. Mikhailov and A.V.Radyushkin, Nucl. Phys.~{\bf B 273}
 (1986) 297.

\bibitem{MMS97} L.~Mankiewicz, M.~Maul and E.~Stein,
Phys. Lett. {\bf B 404} (1997) 345.

\bibitem{V80} A.A. Vladimirov, Theor.Math. Phys.~{\bf 36} (1978) 732;
~{\bf 43} (1980) 417.
\bibitem{FORM} J.A.M. ~Vermaseren, Symbolic Manipulation with FORM,
Version 2, Tutorial and Reference Manual (Computer Algebra Nederland,
Amsterdam,1991).
\bibitem{Leo}  L. V.~Avdeev et. al., Phys. Lett. {\bf B 336} (1994) 560;
 L. V.~Avdeev, Comp.Phys.Commun. ~{\bf 98} (1996) 15;
L.V.~Avdeev, J.~Fleischer, M. Yu. ~Kalmykov, M.~Tentyukov,
{\em ``Towards automatic analytic evaluation of massive Feynman diagrams"},
~hep-ph/9610467.
\bibitem{DMRGH88}F.M. Dittes, D. M\"uller, D. Robaschik, B. Geyer and
  J. Horejsi, Phys. Lett.~{\bf B 209} (1988) 325.
\bibitem{Mak81} Yu. M. Makeenko, Sov. J. Nucl. Phys. {\bf 33} (1981) 440.
\bibitem{GK97} P. Gosdzinsky and N. Kivel,
{\em ``Resummation of $(-b_0 \alpha_s)^n$ corrections to the photon-meson
transition form factor  $\gamma^* + \gamma \to \pi^0$''},
 hep-ph/9707367.
\end{thebibliography}
\end{document}